\documentclass[jgrga]{agu2001}
\usepackage[dvips]{graphicx}

\authoraddr{Agn\`es Helmstetter, LGIT, Maison des G\'eosciences, BP 53, 38041 Grenoble Cedex 9 , France
 (e-mail: ahelmste@obs.ujf-grenoble.fr)}
\authoraddr{Bruce Shaw, Lamont-Doherty Earth Observatory,
61 Rte 9W, Palisades, NY 10964 (e-mail: shaw@ldeo.columbia.edu)}

\newcommand \be{\begin{equation}}
\newcommand \ba{\begin{eqnarray}}
\newcommand \bas{\begin{eqnarray*}}
\newcommand \ee{\end{equation}}
\newcommand \ea{\end{eqnarray}}
\newcommand \eas{\end{eqnarray*}}
\lefthead{Helmstetter and Shaw }
\righthead{Afterslip, aftershocks and rate-and-state friction}
\begin{document}

\setkeys{Gin}{draft=false} 

\def\today{\ifcase\month\or
 January\or February\or March\or April\or May\or June\or
 July\or August\or September\or October\or November\or
December\fi
 \space\number\day, \number\year}

\title{Afterslip and aftershocks in the rate-and-state friction law}


\author{Agn\`es Helmstetter$^1$ and Bruce E. Shaw$^2$ }
\affil{$^1$ Laboratoire de G\'eophysique Interne et Tectonophysique, Universit\'e 
Joseph Fourier and Centre National de la Recherche Scientifique, Grenoble, France \\
$^2$ Lamont-Doherty Earth Observatory, Columbia University, New York }

\begin{abstract}

 We study how a stress perturbation generated by a mainshock affects 
a population of faults obeying a rate-state friction law.
Depending on the model parameters and on the initial state, the fault exhibits 
aftershocks, slow earthquakes, or decaying afterslip.
We found  several regimes with slip rate decaying as a power-law of time, with different characteristic times and exponents.
 The complexity of the model makes it unrealistic to invert for the friction law parameters from afterslip data. We modeled afterslip measurements for the Southern California Superstition Hills earthquake using the complete rate-and-state law, and  found a huge variety of model parameters that can fit the observed data. In particular, it is 
impossible to distinguish  the stable velocity strengthening regime ($A>B$) from the (potentially) unstable velocity weakening regime ($B>A$ and stiffness $k<k_c$).
Therefore, it is not necessary to involve small scale spatial or temporal 
fluctuations of friction parameters $A$ or $B$ in order to explain the transition between stable sliding and seismic slip. In addition to  $B/A$ and stiffness $k/k_c$, 
the fault behavior is strongly controlled by  stress levels following an event.  Stress heterogeneity can thus explain 
most of the variety of postseismic behavior observed in nature.
Afterslip will induce a progressive reloading of  faults that are not slipping, which can trigger aftershocks.
Using the relation between stress and seismicity derived from the rate-and-state friction law, 
we estimate the aftershock rate triggered by afterslip.
Aftershock rate does not simply scale with stress rate, but exhibits different characteristic times and sometimes a different power-law exponent.
 Afterslip is thus a possible candidate to explain  observations of aftershock rate decaying as 
 a power-law of time with an Omori exponent that can be either smaller or larger than 1.

\end{abstract}

\begin{article}

\section{Introduction}

Most shallow large earthquakes are followed by a significant postseismic deformation, and 
by an increase in seismic activity, which both start immediately after the mainshock and last for 
several years. 
The link between aseismic afterslip and aftershock activity is however not clear.
The cumulative moment released by aftershocks is usually much lower than the one 
associated with afterslip, which implies that postseismic deformation is unlikely to be 
due to aftershock activity.
The similar time decay and duration of postseismic deformation and aftershocks 
rather suggests that aftershocks are induced by afterslip \citep{wennerberg97,schaffbs98,perfettini04,hsu06}.
But there are alternative models which explain aftershock triggering by the static 
 \citep{dieterich94} or dynamic  \citep{gomberg98} stress change associated with the mainshock, or by fluid flow  \citep{nur72}.

Postseismic deformation is most often localized around the rupture zone, and is thus modeled
as afterslip on the mainshock fault. 
However,  there are observations of long-range diffuse deformation following 
large earthquakes \citep{nur74}, which can be modeled by visco-eslatic relaxation of the lower crust 
or upper mantle.  Another potential candidate for postseimic deformation is poro-elastic deformation.
Distinguishing between the different mechanisms is difficult based on available data \citep{montesi04}. 
In some cases, several processes have to be involved to fit the data \citep{pollitz06,freed06}.
Afterslip is often comparable with postseismic slip (see   \citep{pritchard06} for a review on afterslip in 
subduction zones), even if there are very large variations in the amount of afterslip from one event 
to another one \citep{melbourne02,marone98,pritchard06}. For instance, the $m_w7.6$ 1994 
Sanriku-Haruka-Oki earthquake in Japan had very large afterslip, with cumulative seismic moment 
a little larger than the coseismic moment \citep{heki97}.
 \citet{takai99} also reported afterslip as large, in term of seismic moment, as coseismic 
 slip for a much smaller $m_w=5.7$ earthquake in the same area. This suggest that 
 afterslip scales with coseismic slip. 

The rate-and-state friction law, introduced by  \citet{dieterich79} based on laboratory friction experiments, 
has been frequently used to model both  aseismic deformation 
\citep[e.g.,][]{marone91}, slow earthquakes \citep[e.g.,][]{yoshida03},  and seismic activity \citep[e.g.,][]{dieterich94}. 
Depending on the parameters of the rate-and-state friction law, the model is either stable (aseismic slip), 
or produce slip instabilities (earthquakes).  
In the unstable regime, the rate-and-state friction law provides a relation between stress history and 
seismicity  \citep{dieterich94}. 
This relation can be used to predict the seismicity rate 
triggered by any stress change, such as static \citep{dieterich94} or dynamic  \citep{gomberg98}  stress change induced
 by a mainshock, postseismic slip, or transient deformation associated with intrusions or eruptions \citep{dieterich00},
 slow earthquakes \citep{ segall06} or tides \citep{cochran04}. 
 Extensions of the original theory to include stress heterogeneity as a 
fundamental aspect of aftershock process has further improved matches
with observations, including 
where aftershocks occur and modifications
to the time dependence of the decay \citep{marsan06,helmstettershaw06}.
This extension explains why many aftershocks 
occur on the mainshock rupture area, where stress decreases on average after the mainshock.

Previous studies have modeled afterslip using the rate-and-state friction law 
\citep{scholz90,marone91,marone98,wennerberg97,schaffbs98,miyazaki04,montesi04},
or a simpler rate-dependent friction law  \citep{perfettini04,hsu06}.
These studies have assumed that afterslip is associated with stable faults in the velocity strengthening regime.
They have also assumed that faults are close to the steady state regime during afterslip, which amounts to neglect of fault healing. With this approximation, the friction coefficient only depends on slip velocity, which simplifies the analysis.
These studies have been very successful in matching afterslip data 
 \citep{scholz90,marone91,marone98,wennerberg97,miyazaki04,montesi04}. 
However, they also bring new problems.
The observation that afterslip  and earthquakes often occur on the same parts of a fault 
requires either small-scale spatial  \citep{miyazaki04} or temporal \citep{wennerberg97} 
variations in the friction parameters.

In this work, we study analytically and numerically the 
postseismic slip in the rate-and-state model, without using the steady state approximation.
We show that afterslip is not limited  to stable faults, but can also occur in seismogenic zones. 
We also fit the model to the afterslip data measured following the 1987 Superstition Hills earthquake  
and analysed by  \citet{wennerberg97}.

Afterslip transfers stress from sliding to locked part of the fault, and is thus a potential mechanism for aftershock triggering
\citep{wennerberg97,schaffbs98,perfettini04,hsu06}. 
\citet{dieterich94} noted that non-constant stressing following an earthquake can alter the aftershock decay rate, and give rise to Omori exponents larger than 1.
The  suggestion that afterslip triggers aftershocks  is based on the observation that both afterslip and aftershock rate roughly decay as the inverse of the time since the mainshock. 
However, there is a priori no  reason to expect that aftershock rate is proportional to stress rate.
 In particular, in the rate-and-state friction model, earthquakes can be triggered at very long times following a stress change.
 The relation between stress history and seismicity rate is indeed complex and non-linear \citep{dieterich94}.
We thus use the complete rate-and-state friction law in order to model afterslip and aftershock activity.
This allow us to better explain the temporal and spatial distribution of afterslip and aftershocks.

\section{Rate-and-state friction law and afterslip}

\subsection{Slider-block model with rate-and-state friction}

We use the rate and state friction formulation   of  \citet{ruina83}, based on \citep{dieterich79}
\be
\mu =\mu^* +A\ln {V \over V^*} +B\ln { \theta V^*  \over D_c} ~~.
\label{ratestate}
\ee
where
$\mu^*$  and $V^*$ are constants,
$A$ and $B$ are friction parameters,   $D_c$ is the critical slip distance, and $\theta$ is the state variable,
which will evolve with slip and time. The state variable is often interpreted as the average age of contacts 
on the fault.

There are a couple of evolution laws for the state variable.
\citet{dieterich94} used the slowness law
\be
{d\theta\over dt}=1-{V\theta\over D_c}, 
\label{ratestate2}
\ee

Following \citet{dieterich94}, we model a fault by a slider-spring system. 
The friction coefficient of the block is given by 
\be
\mu= {\tau \over \sigma}=  {\tau_r  -k \delta\over \sigma}
\label{mu}
\ee
where $k$ is the spring stiffness, $\sigma$ is the normal stress, 
$\tau$ the shear stress on the interface, $\tau_r$ is  the remotely aplied stress acting 
on the fault in the absence of slip, and $-k  \delta$ is the decrease in stress due to fault slip.
Expression (\ref{mu}) neglects inertia, and is thus only valid for low slip speed in the interseismic period.
The stiffness  is a function of the crack length $l$ and shear modulus $G$ \citep{dieterich94}
\be
k \approx G / l \,.
\ee

\subsection{Steady state approximation}

In the steady state regime $\dot \theta=0$, the friction law (\ref{ratestate}) becomes
\be
\mu_{ss}(V) =\mu^ * + (A-B)  \ln {V \over V^*}  ~~.
\label{ss}
\ee
If $A>B$, the friction law is velocity-strengthening, and the system is stable.
In this regime, the state variable relaxes toward its steady state value when the slip exceeds the critical slip $D_c$, 
and at constant sliding velocity.
If $A<B$, the steady state (\ref{ss}) is unstable, except during inertia controlled instabilities \citep{rice86}, and 
unless the system is too stiff. 
\citet{ruina83} showed that, in order to produce slip instabilities ("earthquakes"), we need both $B>A$ 
and $k<k_c$, where $k_c$ is defined by 
\be
k_c={ \sigma  (B-A) \over D_c}\,. 
\label{kc}
\ee
Note that $k_c$ defined by (\ref{kc}) is negative for $A>B$.

 In  the steady state approximation,  the state variable  is constant, there is thus no healing. 
 This approximation  $\dot \theta=0$ in (\ref{ratestate2}) also requires that the slip speed is approximately constant equal to $D_c/\theta$.
 Therefore, it does not seem relevant to model afterslip, when slip speed 
 decreases between coseismic slip speed ($\approx m/s$) at very short times, to values lower than the long-term rate
  ($\approx mm/yr  \approx 10^{-10} m/s$) at large times.

Previous studies \citep{scholz90,marone91,wennerberg97,marone98,schaffbs98,montesi04,hsu06} 
have used the rate-and-state friction law to model afterslip, assuming the fault is close to the steady state regime.
They suggested that afterslip is produced within velocity-strengthening ($A>B$) parts of the fault, below or above 
the seismogenic zone (where $B>A$ to produce earthquakes).
In these zones, there is a slip deficit after an earthquake, and thus an increase in Coulomb stress, which is 
relaxed during afterslip. 

With the steady state approximation $\dot \theta\approx 0$, and without loading stress rate ($\dot \tau_r =0$),  afterslip increases logarithmically with time  \citep{scholz90,marone91} 
\be
\delta = {(A-B) \sigma \over k }  \ln \left(   {  k V_0 t \over \sigma (A-B)}  +1  \right) 
 = V_0 t^* \ln \left(   { t \over t^*  } +1\right) 
\label{slipss}
\ee
where $V_0$ is the initial slip speed and 
\be
t^*=  {\sigma (A-B) \over k V_0}
\label{ts}
\ee
 is a characteristic time for afterslip.
Slip speed given by
\be 
V = { V_0  \over  1+t/t^*} 
\label{vss}
\ee 
 decreases as the inverse of time  for $t>t^*$.

 We can test the limit of validity of equation  (\ref{slipss}) by injecting the solution for the slip 
 speed (\ref{vss}) and the state variable $\theta=D_c/V_0$ back into the evolution law (\ref{ratestate2})
 \be
 \dot \theta = 1- {V \over V_0}  = 1 -  {1 \over 1 + t/t^*} 
 \ee
 This shows that the steady state approximation (\ref{slipss})  works only for very short times, when velocity 
 is close to its initial value, but gets very bad when $t>t^*$. Thus it cannot explain 
 the power law decay of afterslip rate with time.
  
Nevertheless, application of expression (\ref{slipss}) provides a very good fit to afterslip data, 
with however significant discrepancies \citep{wennerberg97,montesi04}.  To better explain the data,  
\citet{wennerberg97} have used a more complex 
 form of equation (\ref{ratestate}). They have also suggested to introduce a second state variable 
 in  (\ref{ratestate}), in order to explain apparent negative values of $A$ obtained when inverting afterslip data, 
 and to explain the transition between aseismic and seismic slip during the earthquake cycle.
 
 \citet{montesi04} introduced a non-zero loading stress in order to better fit GPS data.
But nobody has yet checked the consistency of the steady-state assumption and verified that 
$\dot \theta $ in the evolution law (\ref{ratestate2}) is indeed negligible during afterslip.
  Only  \citet{marone91} performed numerical simulations without 
  the quasi-static approximation (\ref{ss}), which were in good agreement with the analytical 
  solution (\ref{slipss}) for a few cases.
  
 While most afterslip generally occurs below or above the seismogenic zone \citep{marone91,hsu06},
 there are observations suggesting that afterslip can also occur along strike the rupture zone 
 \citep{melbourne02,pritchard06,chlieh07}, 
 or even within the rupture area  \citep{miyazaki04}, and can be associated with aftershocks \citep{miyazaki04}.
To explain these observations within the steady state approximation, we need to invoke 
small scale spatial or temporal variations of the friction parameters   $A$ or $B$ \citep{miyazaki04,wennerberg97}.
 
  \citet{miyazaki04} attempted to distinguish between velocity-weakening and velocity strengthening behavior 
  for different parts of the fault that slipped after the 2003 Tokachi-oki earthquake. They inverted afterslip and 
  stress changes on the fault from GPS data. They found that, within the rupture area, slip rate and stress initially 
  decrease with time. But then stress increases while slip velocity continues to decrease.
  For other parts of the fault around the rupture area, stress decreases roughly linearly with the log of the slip rate, 
  as expected with the steady state approximation (\ref{ss}).   \citet{miyazaki04}  interpreted these results as evidence for 
velocity-weakening (unstable) behavior within the rupture area, and   velocity-strengthening regime around 
the mainshock rupture.
However, the steady state approximation (\ref{ss}) is only valid at constant slip speed, while slip rate changes by 
several orders of magnitude in their observations. The different behavior between the rupture area 
and surrounding regions may rather arise from different stress histories: stress increases within the rupture area 
because it is reloaded by surrounding region, where afterslip is larger.

  \citet{marone91}  also evoked the possibility that afterslip may be produced by unstable faults, but noted that 
  "for a fault that exhibits only velocity-weakening behavior, the steady-state frictional resistance decreases with slip 
  velocity, eliminating the stress transient needed to drive afterslip." 
  However, it is possible to produce significant afterslip starting from the steady state  with $B>A$, and with 
  realistic values of $D_c$.
 \citet{rice86} studied a simple slider-block model in the unstable regime including inertia, and showed that 
the slider is close to steady state during inertia controlled instability, when slip exceeds $D_c$. During dynamic rupture,  
stress first decreases as slip velocity increases, but further increases as slip decelerates. When inertia becomes 
negligible, the system escapes from the steady-state regime, and both slip rate and stress slowly decrease with time 
(see fig 3a of \citep{rice86}). Thus, even if the slider is in the steady state regime at the end of dynamic rupture, 
there can be significant afterslip for velocity weakening faults.
In addition,  earthquake rupture is more  complex than with single degree of freedom systems.
There may be parts of the fault where slip is smaller than $D_c$, and 
stress transfers during dynamic rupture propagation can produce stress concentration, so that 
stress can be locally larger than its steady state value after a mainshock.
  

In this section, we study analytically and numerically the complete rate-and-state friction law (\ref{ratestate},\ref{ratestate2})
 both in the velocity-strengthening and in the  velocity-weakening regimes.
 We then attempt to invert the friction law parameters by fitting the afterslip data of \citet{wennerberg97}
  for the 1987 Superstition Hills earthquake.
 
\subsection{Stability condition}
 
 Depending on  the parameter $B/A$, stiffness $k/k_c$, and stress, the 
 system exhibits either decaying afterslip, slow earthquakes, or slip singularity ("aftershock").
  \citet{rice86} derived the condition of stability for a slider without loading rate ($\dot \tau_r=0$) with $B>A$ and $k<k_c$.
 They found that the system is stable if $\mu <\mu_l $ with
 \be
 \mu_l (V) = \mu_{ss} (V) + { k D_c B \over (B-A) \sigma }  =  \mu_{ss} (V) + { k B \over k_c }  
 \label{mul}
 \ee
 
We compute below the limit of the friction coefficient above which initial acceleration is positive.
 We rewrite the rate-and-state friction law  (\ref{ratestate}) 
 and (\ref{mu}) as 
 \be
 V = V_0 \,   e^{ - k \delta / A\sigma} \, {\left( \theta \over \theta_0 \right)}^{-B/A}
 \label{V}
 \ee
 where $V_0$ and $\theta_0$ are the initial values of $V$ and $\theta$ respectively.
Taking the time derivative of (\ref{V}), acceleration of the slider is given by
\be
\dot V = V \left (- {k V\over A \sigma}- {B \dot \theta \over A \theta} \right)\,.
 \label{Vdot}
\ee
Thus slip accelerates if $ \dot \theta < - k V \theta / \sigma B$. 
Using the state evolution law (\ref{ratestate2}), we get the limit value for the state rate 
\be
 \dot \theta  = { 1 \over 1 - B \sigma / k  D_c } = { 1 \over 1 -  k_c / (k (1-A/B)) }
 \label{dtha}
 \ee

We can rewrite eqs (\ref{ratestate},\ref{ratestate2},\ref{ss}) to get a relation between friction and state rate
\be
\mu = \mu_{ss} (V) + B \ln ( 1 -  \dot \theta )
 \label{mudth}
 \ee
which gives the value of the friction $\mu_a$ corresponding to  $\dot V =0$
\be
\mu_a(V) = \mu_{ss} (V) + B \ln \left({ 1 \over 1 - (1- A/B) k  / k_c }\right)
\label{mua}
\ee

If both $V$ and $\dot \theta$ increase, the slider will eventually decelerate before reaching slip instability.
For the system to reach instability, we need both  $\dot V  >0$,  $\dot \theta <0$ and $\ddot \theta <0$ 
(this is what we observe in the numerical simulations).
The state acceleration is given by (taking the time derivative of (\ref{ratestate2}), and using expression (\ref{Vdot}) of $\dot V $, and expression (\ref{ratestate2}) of $\dot \theta $)
\be
\ddot \theta ={ - \dot \theta V - \theta \dot V  \over D_c} = 
 {  \dot  \theta V  \over D_c} \left ( {B\over A } -1 -   {k D_c\over A \sigma } \right)+ {k V\over A \sigma } 
\ee
The system is thus unstable if $\ddot \theta < 0$, which gives $\dot \theta  < 1/(1- k_c / k)$
and yields another expression for the friction coefficient $\mu_l$ at the stability limit 
\be
\mu_l(V) = \mu_{ss} (V) + B \ln \left({ 1 \over 1 - k  / k_c }\right)
\label{mul2}
\ee
Slip instabilities can occur  only when both $B>A$ and $k<k_c$; otherwise we can't have 
both $\dot \theta<0$ and $\ddot \theta<0$.
 The friction at the stability limit  (\ref{mul2}) is close to but larger than the condition for initial acceleration (\ref{mua}) of the slider.
Between these two values, there is thus a range of parameters for which we observe "slow earthquakes", followed 
by a classic afterslip relaxation, in both the velocity strengthening or velocity weakening regimes.
This behavior is similar to the observation by \citet{pritchard06} of a slow slip event triggered by 
the 1995 $m_W=8.1$ Chile earthquake.

 \subsection{ Analytical study}
 
We have performed numerical simulations and analytical study of the slider block model 
with a fixed loading point ($\dot \tau_r=0$).
We found different regimes for which the model exhibits afterslip, with 
a slip rate decreasing approximately as a power-law of time.
If initial stress is large enough, and for special values of the model parameters,
the system exhibits slow earthquakes or aftershocks.
The analytical calculations are shown in appendix A, while the main results are presented in Table 1.
Figure \ref{figq} illustrates the different regimes  as a function of $B/A$ and $k/|k_c|$.
The other parameter that controls the behavior of the slider is   initial stress.
The influence of initial stress is illustrated in Figure \ref{figabmu1} for $k/|k_c|=0.8$ and    in Figure 
\ref{figabmu2}  for $k/|k_c|=2.5$. 
Figure \ref{figV} shows the temporal evolution of velocity, state variable, and friction, 
for numerical simulations with (top plots) $B/A=1.5$ and $k=0.8k_c$, and (bottom plots) 
$B/A=0.5$ and $k=2.5|k_c|$. Each curve corresponds to a different value of initial friction.

 \subsubsection{ Slip and slip rate}
 We have found several solutions for the slip rate of the form  
 \be
  V= { V_0 \over (1+t/ t^*)^p} \,,
  \label{Vp}
 \ee
  where the exponent $p$ is either smaller, equal or greater than 1 depending on the model parameters, and 
 $t^*$ is a characteristic time, which depends on the initial conditions and on the friction law parameters.
 Generally, the system evolves from one regime to another one with time, so that the apparent exponent 
 $p=d \ln V/d \ln t$  can either increase or decrease with time.

In all cases with slip rate decaying with time,  displacement is proportional to $D_c$. 
Depending on model parameters $B/A$, $k/|k_c|$, and initial 
friction, total slip can be either much smaller or larger than $D_c$.
In the first regime, as  $\delta$ becomes of the order of $D_c$, 
the system evolves from regime \#1 to \#4 if $A>B$, or from  \#1 to \#6 if $A<B$.
The  regimes \#5 and \#6 require  slip to be much smaller than $A\sigma/k$. If $A>B$, slip will eventually become larger
than $A\sigma/k$, and the system will evolve from regime \#5 to \#4.

If $B>A$, slip in regime \#6 increases much slower with time, and reaches a constant value at infinite times. 
There is thus no transition to another regime.
If  $B\approx A$ and $ \dot\theta_0 \ll 1$, then total afterslip on seismogenic faults (with $k<k_c$) can be much larger than $D_c$. 

When there is a transition to another regime, analytical solutions in Table 1 have to be modified by replacing 
the initial values $\theta_0$, $V_0$, and $\mu_0$ by their values at the time the system enters the new regime, and the 
slip produced previously has to be added to the solution for $\delta$ and $\mu$.

In the unstable regime, for $k<k_c$ and $B>A$, the system will eventually 
reach regime \#6 if $\mu_0 < \mu_{l}$, otherwise it produces a slip instability (aftershock).
 
 \subsubsection{ Characteristic times}
In all cases, characteristic times $t^*$ for afterslip are inversely proportional 
to initial slip rate: the faster it slips, the faster slip rate decays with time. 
The other parameters that controls $t^*$ are friction parameters $B/A$, stiffness $k/|k_c|$, 
$D_c$, and, for cases \#5 and \#6, the initial friction or state rate. 

\subsubsection{ State and state rate}
In the velocity strengthening regime $A>B$, state rate 
evolves toward $\dot \theta _l = 1/(1-k_c/k)$.
Because $k_c<0$ when $A>B$, the asymptotic value of the state rate is in the range  $0<\dot \theta _l<1$.
Thus, except for the special case $k=k_c$, the system never reaches the steady-state regime $\dot \theta =0$.
Decrease in velocity with time in the evolution law (\ref{ratestate2}) has to be balanced by an increase 
in state and/or state rate. Therefore, the steady state regime $\dot \theta=0$ can be reached only at constant slip rate.
We consider the load point is fixed. Thus slider velocity cannot  evolve toward a positive fixed 
value, and the slider never stays in the  steady state.
In the velocity weakening regime $A<B$, state rate either decreases toward $-\infty$ 
if $\mu_0 > \mu_l (V_0)$, or increases toward 1 if $\mu_0 < \mu_l (V_0)$.
The steady state approximation $\dot \theta \approx 0$ 
is thus valid only for $A>B$ and $ k \ll k_c$.

\subsubsection{Friction}
Friction during afterslip always decreases with time  in the absence of 
an external loading ($\dot \tau _r=0$), because $\mu = \mu_0 - k \delta /\sigma$.
Thus even in the  "velocity strengthening regime", friction decreases with slip rate if 
slip rate decreases with time, while steady-state friction $\mu_{ss}(V)$ increases with $V$. 
Change in friction with $\ln V$ is different among the afterslip regimes in Table 1.
The slope of $\mu$ as a function of $\ln V$ is $A+qB$ in cases \#1-3,  $A-B$ in case \#4 (with $A>B$), while 
friction is almost independent of $\ln V$ in cases \#5 and \#6.

\subsubsection{Slow earthquakes}
\label{sloweq}
If initial friction is larger than the condition for initial acceleration, and lower 
than the stability criteria (or in the stable regime $A>B$), velocity first increases, without reaching 
seismic slip velocities. The slip speed then decreases as a power-law of time,
following the regime \#4 if $A>B$, or \#6 if $A<B$. The maximum slip rate occurs at times 
of the order of $t_1^*$.
If $B>A$, and if initial friction is $\mu_{ss}(V_0) < \mu_0 <\mu_a(V_0)$,
there is a transient phase in which velocity is almost constant, and healing rate $\dot \theta$ increases.
This phase can produce aseismic slip larger than $D_c$, before the system evolves 
to regime \#6 when $\dot \theta \approx 1$, and slip saturates.  

 \subsection{ Comparison with observations}
Several studies have attempted to measure $A$ or $A-B$, and to distinguish between 
the stable and unstable regimes from the evolution of stress with slip rate during afterslip.
\citet{miyazaki04} have mapped the coseismic and postseismic slip produced by 
the  2003 Tokachi-oki earthquake, as well as the change in shear stress on  the fault, 
by inverting GPS time series. Most afterslip occurred around the rupture area, mostly downdip. 
Within these zones, the stress-velocity paths  approximately follow $ d\tau/d \ln(v) =  0.6$ MPa.
They interpret this result as an evidence that the main afterslip regions are velocity strengthening.
Following \citet{marone91}, they assume that stress is equal  to its steady state value, 
and  suggest that  $(A-B)\sigma= 0.6$ MPa. According to our analysis, 
this result would be true if the system was in the afterslip regime \#4, but 
we cannot exclude the first regime in Table 1. In this case the slope $ d\tau/d \ln(v) =  0.6$ 
would be equal to $A+qB$ instead of $A-B$, and we cannot distinguish between the 
velocity weakening or strengthening regimes. We can only exclude the unstable case $B>A$ and $k<k_c$, 
which never produces a linear decrease of $\mu$ with $\ln V$.

 It is difficult to explain with the interpretation  of \citet{miyazaki04} why numerous aftershocks occurred in the areas 
 of large afterslip, because aftershocks cannot nucleate within  zones with $A>B$. 
\citet{miyazaki04} suggest small-scale variations in friction parameters $A$ and $B$ within 
the afterslip areas in  order to explain the spatial distribution of aftershocks.
 
  \citet{miyazaki04} also found significant afterslip within the rupture area, but with less slip 
  than the surrounding zones, of the order of 0.1 m instead of 0.5 m downdip of the rupture zone.
  In this zone, stress decreases a little at short times, and then reloads, because 
  afterslip is larger in the surrounding areas.
 They suggest  that this behavior is similar to that of a  single degree of freedom slider block 
model with a constant  loading rate in the velocity weakening regime.

  \citet{hsu06} apply the same method to the 2005 Nias earthquake, and obtain similar result.
  They use a simplified form of the rate-and-state friction  law (\ref{ratestate}) with $B=0$ \citep{perfettini04}, i.e., 
  neglecting healing. 
They observe extensive afterslip up-dip from the main shock and a lack of substantial overlap 
between seismogenic and aseismic regions.
Aftershock zones correspond to the transition between regions of coseismic and aseismic slip.
In the regions of large afterslip, stress decreases roughly linearly with $\ln(V)$, with a slope 
$d\tau / d\ln(v) \approx 0.2$ MPa at short times, but much smaller for  times larger than 100 days.
In the areas of smaller afterslip, shear stress increases a little with time, because 
these zones are reloaded by surrounding sliping areas.  \citet{hsu06} nevertheless measure $A\sigma$ 
from the slope $d\tau / d\ln(v) \approx -0. 02$ MPa, giving unphysical negative values for $A$.

\subsection{Fitting afterslip data with the rate-and-state model}

We use the afterslip data of the 1987 $m _W6.6$ Superstition Hills earthquake in Southern California  
analyzed by \citet{wennerberg97}, as well as displacement data for a creep event on the same fault.
High-quality horizontal and vertical afterslip data were obtained from repeated surveys of six quadrilaterals 
established on the Superstition Hills fault \citep{sharp89a,sharp89b}. The quads 
were distributed along approximately 20 km of the 
fault, from North to South referred to as N, NC, SC, SSC, FSSC, and S (see Fig 1 of  \citep{wennerberg97}).
For each station, we have 2 measurements of horizontal strike-slip displacement across the fault 
(labeled as "$h,S$" and "$h,W$" in Table 2), and one measure of vertical displacement (indicated by a subscript $v$ in 
 Table 2).
Stations N, NC, SC and S were installed before the earthquake.
The other quadrilaterals SSC and FSSC were established 91 days after the mainshock, 
and at these sites the later data were combined with earlier lower-resolution measurements of offset 
natural features. 
\citet{wennerberg97} also analyzed slip measurements of \citep{bilham92} from a digital 
creepmeter a few hundred meters to the North of the FSSC quad (labeled as "$FSSC_{h,BB}$" in Table 2).
Following \citet{wennerberg97}, we analyze the vertical time series only from the two quads S and SC, 
because vertical motions from the other quads are too small. 
Slip resolution is about  1 mm for afterslip data, and 0.001 mm for the creep event.
But fit residuals of  \citet{wennerberg97} are largely controlled by steps ("creep 
events"), whose amplitude can be as large as 7 mm.

\citet{wennerberg97} modeled the afterslip measurements using an alternative form of the rate-and-state 
 friction law (\ref{ratestate}) 
 \be
 \mu = {  1 +  B \ln ( 1 +  \theta V^* / D_c) \over 1 +  A \ln ( 1 + V/V^* )  }
 \ee
 and assuming steady-state behavior $\dot \theta \approx 0$. 
 This friction law  provided a better fit of afterslip data than the steady-state friction law (\ref{ss}) derived by  \citet{scholz90}.
In particular,  \citet{wennerberg97} found that slip velocity was decaying as a power law of time 
( \ref{Vp}) with an exponent $p=1/(1+B-A)$ \citep{wesson88}. 
This model is thus able to explain why
$p$ values in Table 2 are often larger than $1$.
But it requires a very large value of  $A-B =0.29$ in order to reproduce the value $p=1.29$ observed 
for afterslip at point $S_{z,S}$.

  For one dataset, this model provided unphysical negative values for $A$, corresponding 
 to an instantaneous decrease in friction following an increase in slip rate. Such a behavior has never been observed experimentally.
 For the other stations, $A$ is ranging between 0.03 and 0.36, a bit larger than the values $A\approx 0.01$ 
 commonly observed in laboratory experiments \citep{dieterich94}.
\citet{wennerberg97} inverted for 4 model parameters and 1 initial condition.
Comparison of model parameters obtained for  different data sets (2 horizontal slip data $S$ and $W$ at each station, 
or using different time intervals) suggests that the inverted parameters are relatively well constrained. 

In contrast with previous studies \citep{marone91,marone98,wennerberg97,schaffbs98,miyazaki04,montesi04,perfettini04,hsu06}, 
we use the complete rate-and-state friction law instead of the steady-state approximation.
We do not fit directly the displacement data, but the fits to the afterslip data 
obtained by  \citet{wennerberg97} (see equation (18) and Table 1 of   \citep{wennerberg97}), 
which fit the data with residuals of the order of the measurement resolution. 
We invert for 6 parameters (instead of 5 in  \citep{wennerberg97}) : $A$, $B$, $D_c$, $k$, $V_0$ and $\mu_0$. 
We use a Nelder-Mead algorithm to perform the nonlinear minimization of the slip residuals, using a large range 
of initial values for the inverted parameters.
Specifically, the initial model parameters have random values determined by 
$ A =  0.01 \times 10^{2r_A}$,  $ B =  A(1+r_B/2)$, $ V_0=V_{d,0}(1+r_V/2)$, $D_c=0.1 \times 10^{2r_{D_c}}$,
$k= A\sigma \times 10^{r_k}/ D_c$ and $\mu_0=r_\mu$, where $r_A$,  $r_B$, $r_{D_c}$,  $r_k$ and $r_V$ are
chosen according to a gaussian distribution of zero mean and standard deviation equal to 1,  $r_\mu$ is uniformly 
distributed between 0 and 1, and $V_{d,0}$ is the initial velocity determined by  \citet{wennerberg97}. 
For each data set, we have tested more than 100 initial models.
We thus have a good sampling of the range of realistic model parameters.

Observed slip velocity is found to decrease with time approximately as $ V =V_0/(t/t^* + 1)^{p}$
with a characteristic time $t^*$ of the orders of days, and an exponent $p$ ranging between  0.88 and 1.3.
(see values in Table 2). 
Figure \ref{figVtShS} shows the "observed" slip rate (computed from the fit of   \citet{wennerberg97})
and all our fits with the complete rate-state friction law with a residue smaller than 1 mm, for 
the data set $S_{h,s}$ (line 12 in Table 12). A simple fit with $ V =V_0/(t/t^* + 1)^{p}$, i.e., only 3 parameters,
already provides a very good fit to the observations, with a residue of 0.5 mm.
This makes it difficult to invert reliably for the 6 parameters of the rate-and-state law.
Indeed, we find that there is a huge range of model parameters that fit the data within 
the measurement resolution. The inverted parameters appear to be mostly constrained by  our choice of 
initial values in the optimization. Therefore, we did not report the inverted model parameters in Table 2.	
In particular, for most data sets we cannot distinguish between the velocity weakening 
and velocity strengthening regimes, and we can fit the data as well with $k<k_c$ or $k>k_c$.
Only the horizontal slip data of the $N$ quad requires $A>B$. This is the only station 
that has an exponent $p <1$. Such a behavior cannot be reproduced with $A<B$ (except over a limited time interval). 
The $FSSC$ quad produces different results for independent slip measurements at points separated by a few meters. For the creep event, all models that fit the data are in the velocity weakening regime with $k>k_c$.

The critical slip distance inverted from the data ranges from  a few $\mu$m to several km, most
of the time comparable to the initial value used in the optimization.
Total afterslip (of the order tens of centimeter) can thus be much larger or smaller than $D_c$. 
Even for seismogenic faults, with $B>A$ and $k<k_c$, we found models that fitted the data 
with $D_c$ as small as 1\% of the maximum afterslip.

\section{Seismicity rate triggered by afterslip}

\subsection{Relation between stress and seismicity}
Previous studies \citep{dieterich94,schaffbs98} have already suggested 
that afterslip  can trigger aftershocks  by reloading parts of the fault that are locked after the mainshock.  
Most studies who suggested that aftershocks are due to afterslip assumed that 
aftershock rate is simply proportional to stress or strain rate \citep{wennerberg97,schaffbs98,perfettini04,hsu06}.
However, \citet{dieterich94} showed that the relation between seismicity rate and stress rate can be much more complex.
For instance, in the case of a simple stress step $\Delta\tau$ followed by a constant stressing rate $\dot \tau_r$, 
the seismicity rate $R(t)$  is given by  \citep{dieterich94}
\be
R(t)={r \over \left( e^{ - \Delta\tau / A \sigma}-1
\right)~e^{-t/t_a}+1} \,, 
\label{R}
\ee
where $t_a= A \sigma_n /\dot{\tau_r}$ is the duration of the nucleation phase of an earthquake, and is also 
equal to the duration of an aftershock sequence. 
$r$ is the seismicity rate for a constant stressing rate equal to $\dot \tau_r$.
Aftershock rate increases immediately after the stress step, and then decreases with time
 back to $R=r$   for $t\gg t_a$.
Typically, $t_a$ is of the order of years \citep{dieterich94}.
This shows that there can be a very long time delay between a change in stress and 
triggered seismicity. The assumption that aftershock rate is proportional to stress rate
for earthquakes triggered by afterslip thus needs to be verified.

\citet{dieterich94} derived a general relation between stress rate and seismicity rate, which we use 
to model seismicity triggered by afterslip
 \be
R={r\over \gamma \dot{\tau_r}}
\label{g}
\ee
where $\gamma$ is a seismicity
state variable, which evolves as
\be
\dot \gamma ={1\over A\sigma}(1 - \gamma \dot \tau)
\label{gamma}
\ee
We can rewrite (\ref{gamma}) as 
\be
{- A\sigma \dot \gamma  + 1 \over \gamma} =  \dot \tau \,.
\label{gamma2}
\ee
Integrating (\ref{gamma2}) and using the definition (\ref{g}), we get a simple form 
for the relation between seismicity rate $R$, cumulative number of events 
$N=\int_0^t R dt$, and stress change $\tau = \int_0^t \dot\tau dt$
\be
A\sigma \ln \left( {R \over R_0} \right) + { N \dot \tau_r \over r} =\tau \,.
\label{SRN}
\ee

For $R$ changing slowly in time, we can neglect the first term in (\ref{SRN}).
So  the seismicity rate is proportional to the stressing rate, specifically
\be
R={r\dot{\tau}\over \dot{\tau_r}} 
\label{Rs}
\ee

For rapid variations of seismicity rate, we can neglect the second term in (\ref{SRN}).
The seismicity rate thus increases exponentially with stress
\be
R= R_0  \exp( \tau/  A\sigma)
\label{Rf}
\ee 

\citet{beeler03} measured the correlation between seismicity rate and stress in laboratory friction experiments,
with a periodic perturbation of the stress rate superposed to a constant loading rate.
They found that the relation between seismicity rate and stress is in good agreement with 
expression (\ref{Rs}) for slow stress changes, and with equation (\ref{Rf}) for fast stress changes.
Expression (\ref{Rf}) also provides a good fit to seismicity triggered by tides \citep{cochran04}, though 
the small values  of the stress change  $\tau/  A\sigma$ do not allow to test if $R$ increases proportionally 
or exponentially with $\tau$.

\subsection{Application to model seismicity triggered by afterslip}

\subsubsection*{Stress rate}

 We have shown previously that slip rate in the rate and state model
 can be modeled by a power law decay $V= V_0/(1+t/t^*)^p$, 
 with a characteristic time $t^*$, and an exponent $p$ that 
 depend on the friction law parameters $B/A$ and $k/k_c$, and on the initial conditions $V_0$ and $\mu_0$.
 Initial slip rate and friction are likely to be very heterogeneous,  due to variations of coseismic slip.
 As a consequence, areas of the fault that are approximately locked after the mainshock (small $V_0$)
 will be reloaded by adjacent areas with larger slip rate.
Using our simple slider-block model, we can model the stress rate 
induced by afterslip as 
\be 
\dot{\tau} = k V =  {k V_0 \over (1+t/t^*)^p}
\label{taua}
\ee
where $V$ represents the difference in slip rate between "locked" and "slipping" 
parts of the fault.

\subsubsection*{Special case $p=1$}
There is an analytical solution to equation (\ref{gamma}) with stress following (\ref{taua}) 
only for the special case $p=1$, which is given by
\be
R={ r \over  \dot{\tau}_r} \left [  C (1+t/t^*) ^{- n } + {1 + t/t^*  \over \dot{\tau}_0 ( 1   + 1/n) } \right]^{-1} \,,
\label{Ra}
\ee
where $C$ is a constant given by, assuming initial seismicity rate equals $r$, 
\be
C={1 \over \dot{\tau}_r} - { 1 \over   \dot{\tau}_0 ( 1   + 1/n)} \,,
\label{C}
\ee
$n$ is defined by 
\be 
n= \dot{\tau}_0 t^* / A \sigma \,,
\ee 
$\dot{\tau}_0= k V_0$ is initial stress rate. 
For large stress rate, the second term is negligible in (\ref{C}) and $C\approx 1/ \dot{\tau}_r$.

In this case $p=1$, the seismicity rate first increases with time and reaches its maximum value for time equal to
\be
t_{\rm R_{max}} =  t^* \left[   \left ( C \dot \tau_0 (n+1)\right ) ^{ 1/(n+1)} -1 \right] \,.
\label{tmax}
\ee
As stress change increases, $t_{\rm R_{max}}$ decreases.
For times larger than $ t_{\rm R_{max}}$, seismicity rate predicted by (\ref{Ra})
decreases proportionally to the stress rate
\be
R \approx \left( 1   + {1 \over n} \right)   { r \dot \tau \over  \dot{\tau}_r}  \,.
\label{Ra2}
\ee
For large $n$ (large stress rate), expression (\ref{Ra2}) recovers expression (\ref{Rs})
for the stress-seismicity relation that is valid for slowly  varying $R$.

\subsubsection*{General case $p\neq1$}

For $p \neq 1$, we found numerically that expression (\ref{Ra2}) is still a rather good approximation 
of the time  $t_{\rm R_{max}}$ at which $R$ reaches its maximum.
At intermediate  times $t>  t_{\rm R_{max}}$ expression (\ref{Ra2}) also provides a good fit 
to the seismicity rate, i.e., seismicity rate is proportional to stress rate. However, if $p>1$, 
we found in numerical simulations of (\ref{SRN}) that there is a characteristic time $t_c$  
after which expression (\ref{Ra2}) does not hold.
Typical evolutions of seismicity rate and stress rate are shown in Figure \ref{figRStp13} for $p=1.3$ and 
in Figure \ref{figRStp08} for $p=0.8$.

If $p>1$, stress and thus number of events $N$ will saturate for $t \gg t^*$.
The stress change will increase toward a maximum value $ \Delta \tau $ given by 
\be
\Delta \tau = \int_0^\infty \dot\tau dt =  { \dot\tau_0 t^* \over  p-1} \,.
\label{Dtau}
\ee
The first term $\sim \ln(R)$ will become non negligible 
compared with the term $\sim N$ in (\ref{SRN}).   
The seismicity rate for large times $t \gg t^*$ and  $t \gg t_{\rm R_{max}}$ 
will thus be equivalent to that triggered by an instantaneous stress step of amplitude $\Delta \tau$, 
described by equation (\ref{R}).
The "long time seismicity rate", for $ t_c \ll t \ll t_a$, evolves according 
to $ R \approx r A\sigma / t \dot\tau_r$ (assuming $t_c \gg  t_a \exp(-\Delta\tau/A\sigma)$).

The transition from the regime $ R  \approx r  \dot \tau/ \dot\tau_r$ described by (\ref{Ra2}) for 
$ t_{\rm R_{max}} \ll t \ll t_c$ 
to the regime  $ R \approx r A\sigma / t \dot\tau_r$ for $t \gg t_c$ will occur at a time $t_c$ given by 
$ \dot \tau (t_c) =A\sigma / t_c $. 
Assuming $ t_c \gg t^*$, we get 
\be
 t_c = t^* n ^{1/(p-1)} \,.
 \label{tc}
\ee 

If there is a non-zero constant stress rate $\dot\tau_r$, in addition to the stress rate (\ref{taua}) induced by afterslip, 
seismicity rate decays as the inverse of time for $t_c \ll t\ll t_a$, until it reaches its background level $r$.
The number of events triggered by afterslip can be computed directly from (\ref{SRN}).
When $R$ returns to its initial value $R_0=r$ and $\tau =\Delta\tau$, the first term is equal to zero in   (\ref{SRN}) and $N= r \Delta\tau/\dot \tau _r $.
This results is independent of the form of the stress change, as long as 
stress reaches a maximum value  $\Delta\tau$ at long times.
The stress change needed to explain the observed number of aftershocks is thus the same 
for static triggering (stress step of amplitude $\Delta\tau$), and for triggering due to afterslip.

For a slow decay of stress rate with time $p<1$, which requires $A>B$ (see Table 1), 
stress does not saturate for $t\gg t^*$ but instead increases as 
$  \tau =  \int_0^t \dot\tau dt \sim t^{1-p}$ for $t\gg t^*$.
Seismicity rate thus never reaches the regime described by (\ref{R}) with $R \sim 1/t$.
In this case, seismicity rate is proportional to stress rate as predicted by 
(\ref{Ra2}) for $t>t_{\rm R_{max}}$.

This work shows that seismicity rate triggered by afterslip does not always scale 
with slip rate, as assumed by \citep{wennerberg97,schaffbs98,perfettini04,hsu06}.
For instance, we can observe $V \sim \dot \tau \sim t^{-p}$ with $p>1$ but  $R \sim 1/t$.
The characteristic times $t_{\rm R_{max}}$ and $t_c$ that control seismicity rate are also different 
from the characteristic time $t^*$ which controls afterslip rate.


\subsubsection*{Comparison with aftershock data}
Aftershock studies \citep{helmstetter05,peng07} have shown that seismicity rate 
decreases with time according to $R \sim R_0/(1+t/c)^p$, with a characteristic time 
$c$ that is no larger than 10 sec, and an exponent $p$ that can be either smaller or larger than 1, 
and is usually between 0.9 and 1.2  \citep{reasenberg94,helmstetter03}. 
Seismicity rate can increase by a factor up to $10^5$ relative 
to the background rate. The duration of aftershock sequences is of the order of years, i.e.,
about $10^6$ larger than the characteristic time $c$.

In order to explain aftershock triggering by afterslip, we need 
the characteristic times $t^*$ to be of the order of seconds. This is much smaller than 
the values of $t ^* $ of a few days estimated for afterslip data following Superstition Hills earthquake 
\citep{wennerberg97} (see Table 2).
However, afterslip was measured at the surface, we can expect strong fluctuations of $t^*$ with 
depth due to changes of friction parameters or initial values. $t^*$ may thus be much smaller at depth, where aftershocks nucleate.

Afterslip following the $m_W=8.7$ 2005 Nias earthquake, and measured by GPS, 
 also produces $t^*$ of the order of 3 days \citep{hsu06} for all GPS points.
 \citet{hsu06} found that both displacement and aftershock number increase logarithmically with time 
$\sim \ln (1+t/t^*)$, with $t^* \approx 3$ days. However, they counted the number of $m>3$ aftershocks triggered by the mainshock, while the catalog is not complete for such small events, and the completeness level decreases with time after the mainshock. Using only $m\geq5.5 $ aftershocks, we 
measured $t ^* \approx 0.1$ day, much smaller than the characteristic time of afterslip inferred from GPS measurements.
It is thus unlikely that early aftershocks of the Nias earthquake are governed by afterslip.

As shown above, the stress change involved to trigger $N$ events is the same 
for static triggering and for triggering by afterslip. Because the amplitude of 
afterslip is often comparable to coseismic slip \citep{pritchard06}, the number of events triggered by afterslip 
should be comparable to that triggered by coseismic stress change, in the unstable areas of the faults with $B>A$ and $k<k_c$.
If $p \approx 1$, or for static triggering, the stress change needed to produce an increase 
of seismicity rate by a factor $10^5$ is 
$\Delta \tau = A \sigma \ln(10^5)=  11.5 A \sigma$.
If $A\sigma \approx 0.1$ MPa, as suggested by \citet{cochran04}, this gives realistic values, smaller 
than the average stress drop.
On the other hand, if we use the laboratory value   $A\approx 0.01$ \citep{dieterich94},  and 
a normal stress $\sigma=100$ MPa (of the order of the lithostatic pressure at a depth of about  5 km), 
then the stress change needed is $\Delta \tau = 11.5$ MPa. This seems quite large, however 
it is possible that maximum stress change can locally reach such values.
As shown by \citet{helmstettershaw06}, seismicity rate triggered by a heterogeneous stress change 
at short times is controlled by the maximum stress change rather than the mean value.
 
If we want to reproduce observations of seismicity rate decaying with $p>1$ over several 
decades in time, we need even larger stress change than for $p \leq 1$. 
For instance, following the 1992 $m_W7.3$ Landers earthquake, the seismicity rate decreased with time as $R\sim t^ {-1.2 }$  over 5 decades in time. 
To explain this pattern, we thus need $t_c \geq 10 ^5 t_{\rm R_{max}}$. 
We have computed the seismicity rate triggered by the stress change modeled by (\ref{taua}) with $p=1.2$,
using numerical integration of equation  (\ref{SRN}).
We found that the stress change $\Delta \tau$ defined by (\ref{Dtau}) required to get 
$t_c \geq 10 ^5 t_{\rm R_{max}}$ is $ \Delta \tau \approx 50 A\sigma$.

\section{Conclusion}
We have modeled the postseismic slip of faults using the rate-and-state friction law.
The postseismic behavior of faults is more complex than predicted previously 
based on the steady state approximation of the friction law.
We found that, depending on the model parameters $B/A$ and $k/|k_c|$, and on initial friction, the fault exhibits either decaying afterslip, slow earthquakes, or aftershocks.
Afterslip, with slip amplitude comparable or even larger than $D_c$, can be obtained 
for any value of the friction law parameters, even for velocity weakening faults ($B>A$).
We have derived several approximate solutions to describe the evolution of slip, state and friction with time. 
There are several regimes with different characteristic times $t ^*$ and exponents $p$ for which 
afterslip decays according to $V= V_0/(1+t / t^*)^p$. This expression of the slip rate 
provides a very good fit to afterslip data \citep{wennerberg97}, with $t^*$ of the order of days, and $p$ ranging between 0.9 and 1.3. 
The number of parameters of the rate-and-state model is too large in order  to invert from afterslip data.
For most data sets we studied, afterslip can be modeled as well with 
velocity weakening or strengthening friction parameters. 
Therefore, we don't need to involve spatial or temporal changes in the friction law parameter $B/A$ in order to explain the evolution between aseismic and seismic slip. 

Aftershock decay with time (Omori law) is similar to that observed for afterslip rate.
This similarity led several authors to suggest that aftershocks are induced by the postseismic 
reloading of the fault due to afterslip.
Using the relation between stress and seismicity derived by \citet{dieterich94}, we have 
shown that afterslip is indeed a possible mechanism to explain aftershock triggering.
But the relation between slip rate and seismicity rate is more complex than previously thought \citep{wennerberg97,schaffbs98,perfettini04,hsu06}.
The process of earthquake nucleation indeed introduces a time delay between stress change and 
triggered earthquakes. As a consequence, seismicity rate is characterized by exponents and characteristic times that can be different from those that control stress rate.
The complexity of the friction law thus makes it difficult to infer the mechanisms responsible 
for earthquake triggering based on observations of stress changes.
Moreover, we have simplified the problem by considering  uniform values 
of the model parameters, and of the slip rate, by modeling the fault with a simple slider block with one degree of freedom.
We have also neglected other processes that may play an important role in the evolution of faults, such as fluid flow, viscous deformation, dynamic stress changes, and subcritical crack growth, among other mechanisms, all of which could
have their own time dependence. 
The modeling of fault slip and seismicity, and even more the characterization of the fault rheology 
based on seismicity or geodesy data, is thus a difficult challenge,
in terms of finding which mechanism may be causing an observed time dependence.
In this paper, we have added to this difficulty by demonstrating additional
time dependent behavior in the primary candidate, the 
rate and state friction law.

\begin{acknowledgments}
Part of this work was  done while the authors
were at the KITP in Santa Barbara.
This research was supported in part by the National Science Foundation under
grants  PHY99-0794 
and EAR03-37226, and the Southern California Earthquake Center, and by 
European Commission under grant TRIGS-043251.

\end{acknowledgments}
\appendix
\section{Analytical study of afterslip}

We first search for solutions to the rate-and-state friction law with slip rate decaying as the inverse of time
\be
V(t) = { V_0 \over (1+t/t^*)} \,.
\label{Vpl}
\ee
Putting this solution into the rate-and-state law (\ref{ratestate}), we get
\be
 - {k  V_0 t^* \over \sigma} \ln \left(1+ {t\over t^*} \right) =  - A   \ln  \left(1+{t\over t^*} \right) + B  \ln \theta  + \mbox {constant} \,.
\label{mupl}
\ee
The evolution law  (\ref{ratestate2}) becomes
\be
\dot \theta = 1 - {V_0 \over D_c} { \theta \over (1+t/t^*)},
\label{dtpl}
\ee 
whose solution is
\be
 \theta =   K (t^* + t) ^{-q} + { t^* +t \over q +1 }  \,,
\label{thpl}
\ee
where $K$ is a constant and $q= V_0 t^* /D_c$.
Depending on time and model parameters, the first or the second term will dominate in (\ref{thpl}).

\begin{figure}
\centerline{
\includegraphics[width=1\textwidth]{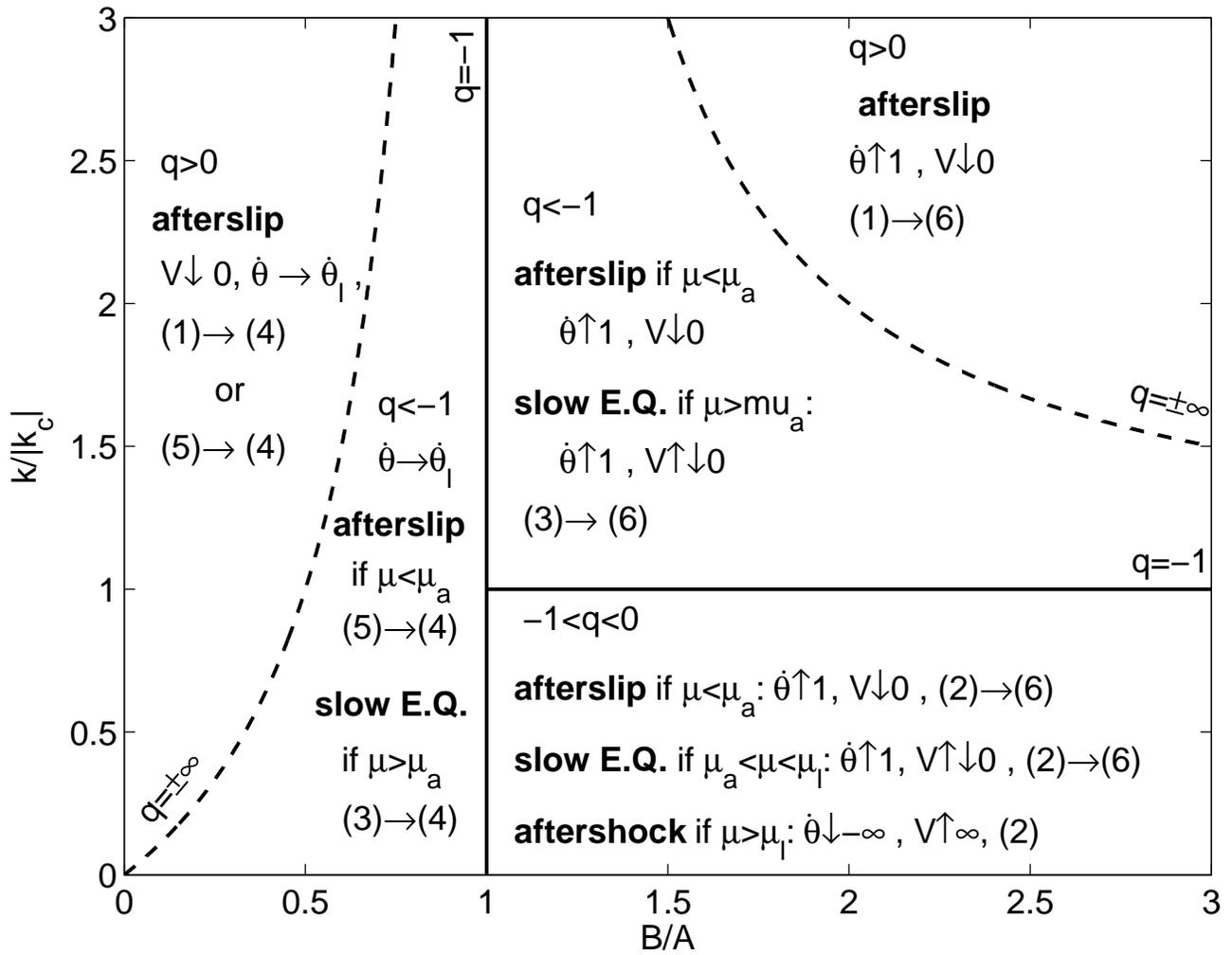} }
\caption{\label{figq}
Value of $q = 1/[(B/A-1) k/k_c -B/A]$ , which controls the different postseismic behaviors (see Table 1).  
}

\end{figure}

\begin{figure}
\centerline{
\includegraphics[width=1\textwidth]{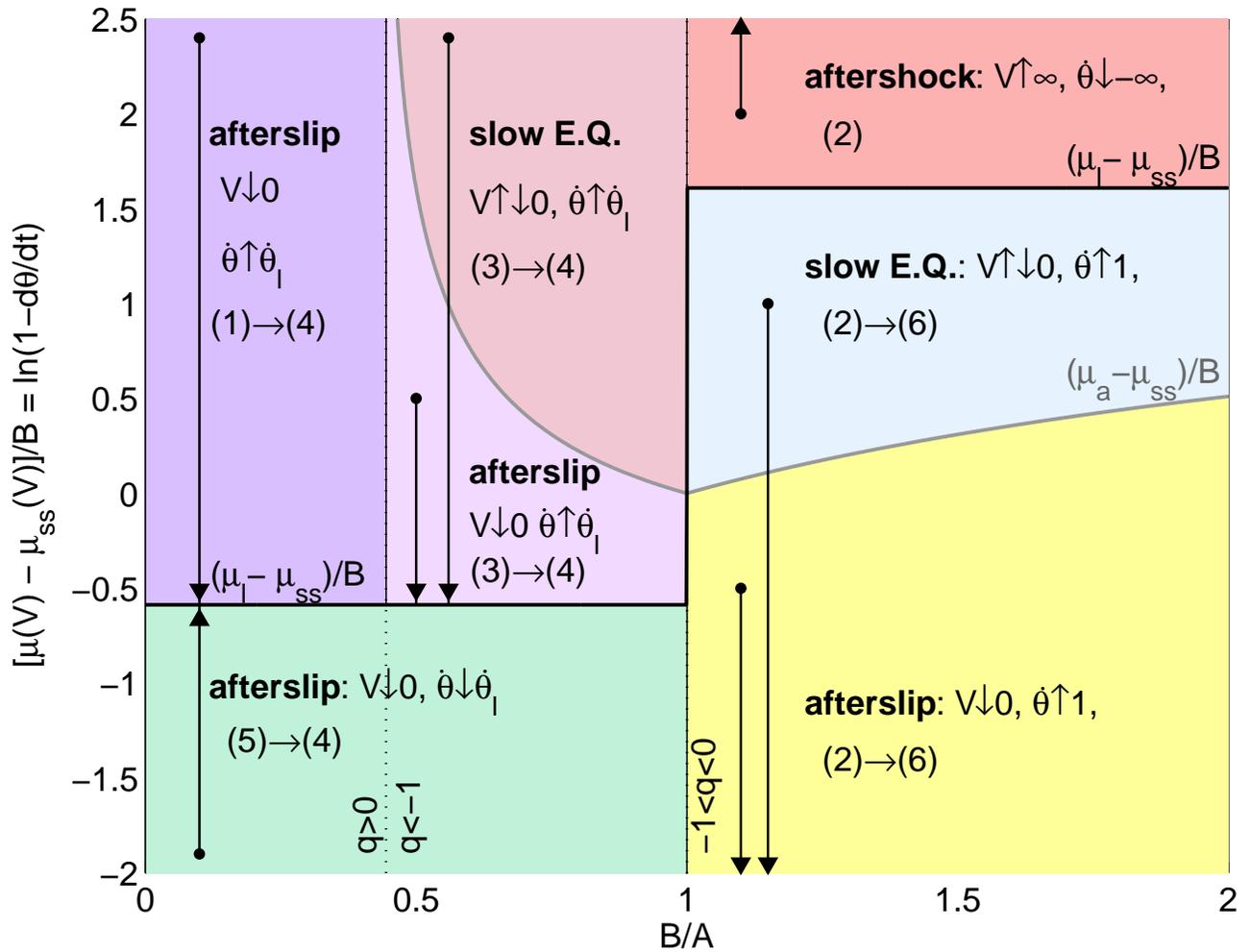} }
\caption{\label{figabmu1}
Value of $\ln(1- \dot\theta)$ as a function of $B/A$ for $k=0.8|k_c|$.  
Also shown are limits for acceleration and instability, and trajectories in this space (arrow).
Numbers in brackets refer to approximate analytical solutions given in Table 1. 
}
\end{figure}

\begin{figure}
\centerline{
\includegraphics[width=1\textwidth]{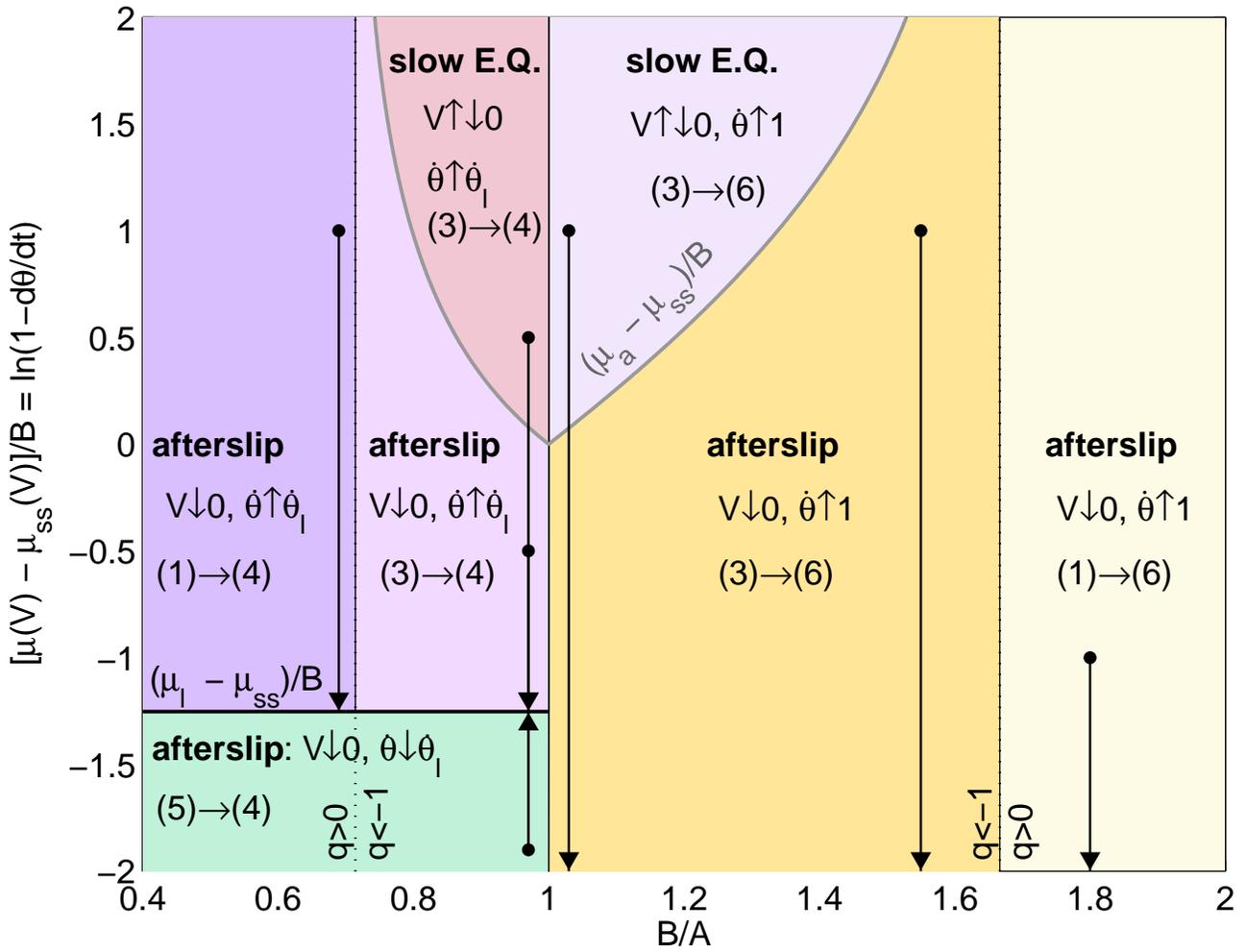} }

\caption{\label{figabmu2}
Same as in Figure  \ref{figabmu1} for $k=2.5|k_c|$.  }
\end{figure}

\begin{figure}
\centerline{
\includegraphics[width=1\textwidth]{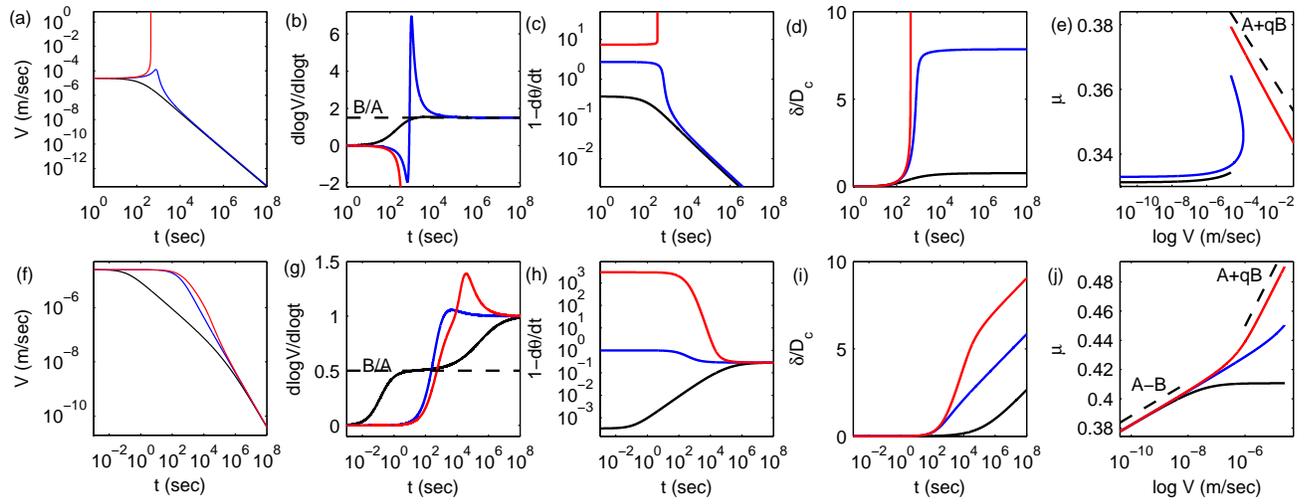} }
\caption{\label{figV}
Temporal evolution of slip rate (a,f),  slip rate exponent $d \ln V /d \ln t$ (b,g), 
state rate (c,h), slip (d,i), as well as variation of friction with slip rate 
(e,j), for numerical simulations with $B=1.5A$ and $k=0.8k_c$ (top, a-e),
and  $B=0.5A$ and $k=2.5|k_c|$ (bottom, f-j).
Each curve corresponds to a different value of initial friction.
In (a-e), initial friction  $\mu_0$ is $\mu_{ss}(V_0)-B$ (black curve),  $\mu_{ss}(V_0)+B$ (blue), and   $\mu_{ss}(V_0)+2B$ (red).
In  (f-j),    $\mu_0$ equals $\mu_{ss}(V_0)-8B$ (black),  $\mu_{ss}(V_0)$ (blue), and   $\mu_{ss}(V_0)+8B$ (red).}

\end{figure}

\begin{figure}
\centerline{
\includegraphics[width=1\textwidth]{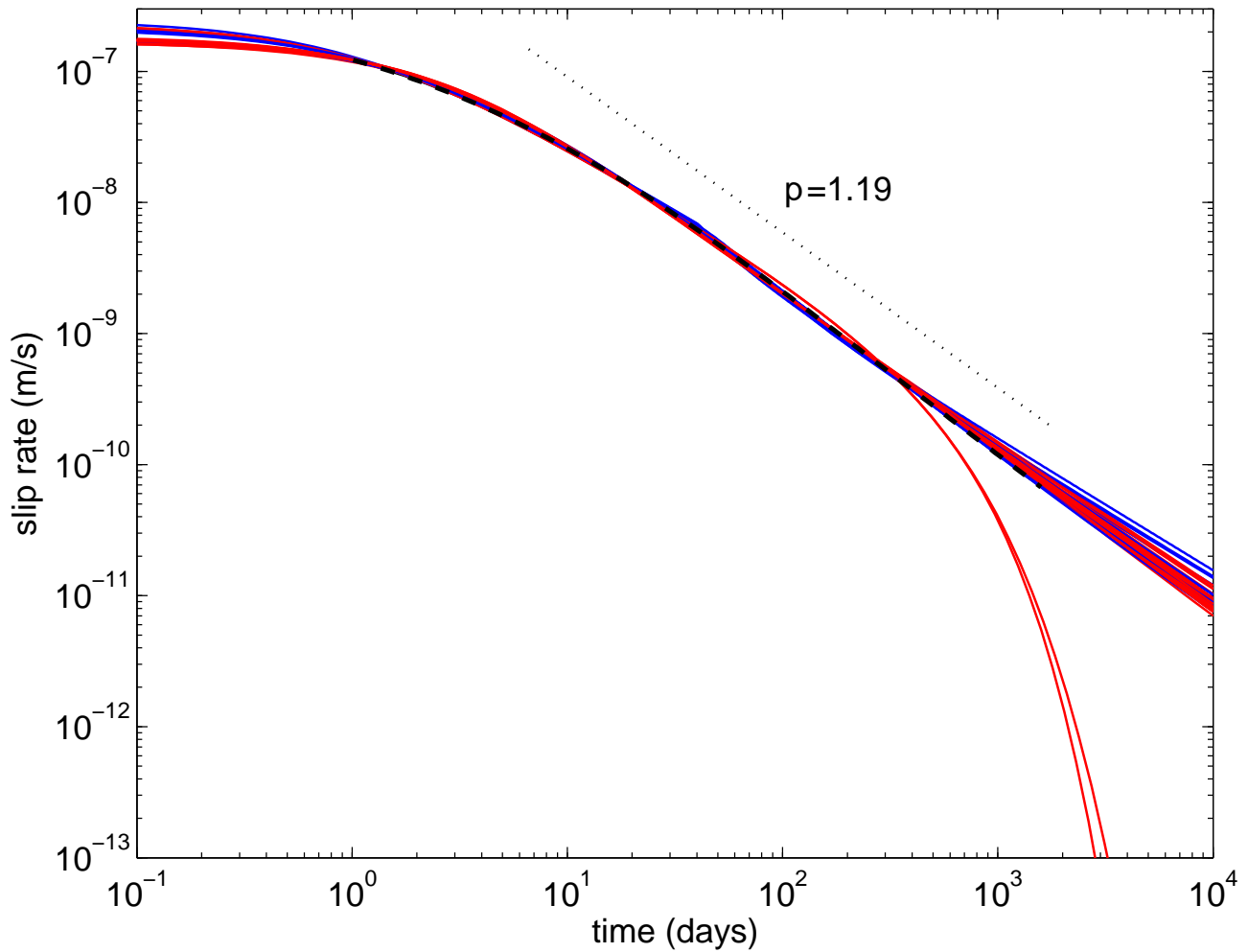} }
\caption{\label{figVtShS}
Observed slip rate for the $S_{h,S}$ time series (see line 12 in Table 2) (dashed black line, for times between 1 and 1600 days),
compared with all models that have a residue (for displacement) smaller than 1 mm.
Models in the velocity weakening regime are shown in red, and blue for 
the velocity strengthening models. The dotted line is the slope measured for $t>6$ days.}
\end{figure}

\begin{figure}
\centerline{
\includegraphics[width=1\textwidth]{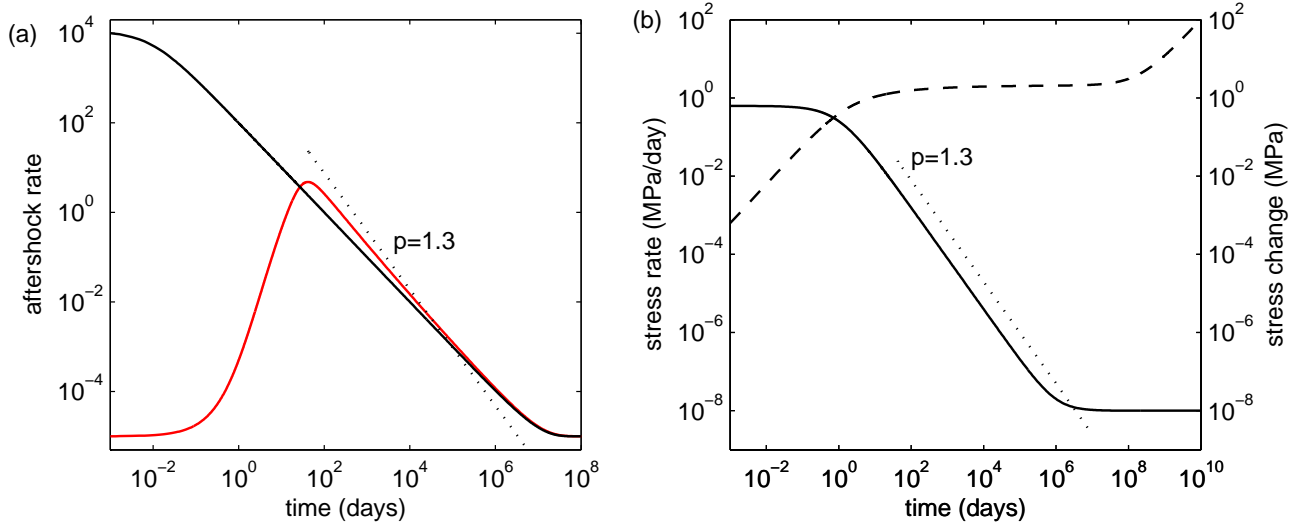} }
\caption{\label{figRStp13}
Seismicity rate (a) triggered by a continuous stress change given by $\dot \tau = \dot \tau_0/(1+t/t^*)^p$ 
(red curve) and by a simple stress step of the same amplitude $\Delta \tau =  \dot \tau_0 t^*/(p-1) =2.08$ MPa
(black curve). Plot (b) shows the stress rate (continuous line) and stress change (dashed line).
The model parameters are $\dot \tau_0 = 0.62$ MPa/day, $t^*=1$ day,  $\dot\tau_r=10^{-8}$ MPa/day 
(constant stress rate) and $A\sigma=0.1$ MPa. This gives $n=6.2$, $C=10^8$.
The time $t_{\rm R_{max}}$ calculated using the approximate solution (\ref{tmax}) is 
 $14.6 $ days, while the observed time when $R$ is maximum is 42 days.
 The crossover time $t_c$ given by (\ref{tc}) is 450 days, and marks the transition between a $\sim 1/t^p$ decay 
 of seismicity rate for $t_{\rm R_{max}}<t<t_c$ to the long time $\sim 1/t$ decay.}
\end{figure}

\begin{figure}
\centerline{
\includegraphics[width=1\textwidth]{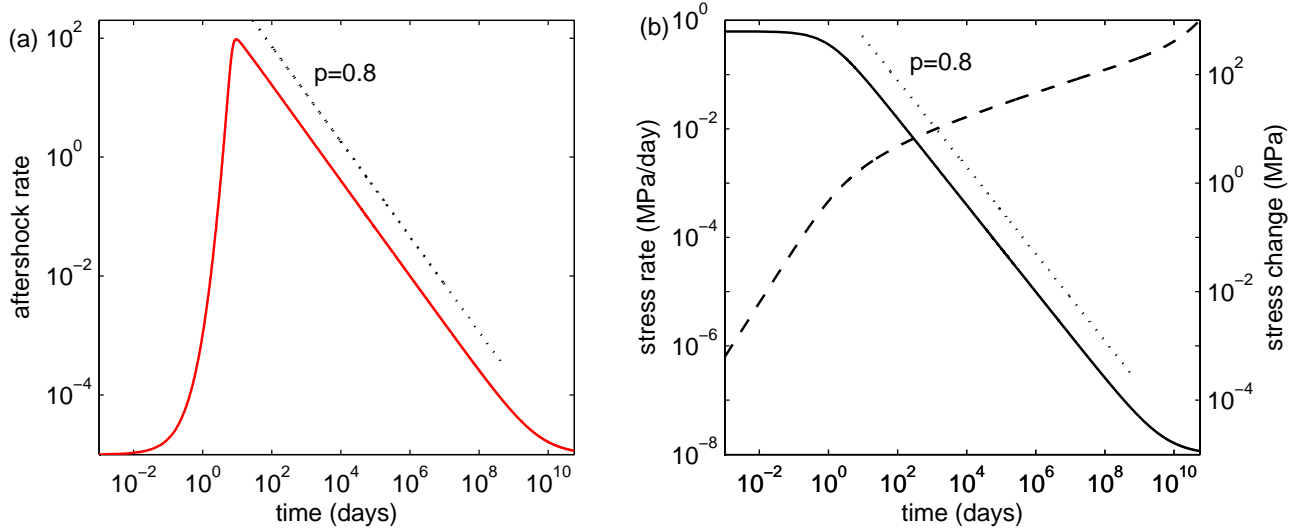} }
 \caption{\label{figRStp08}
 Same as in Figure \ref{figRStp13} for $p=0.8$, but with the same values of $t^*$ and $\dot \tau_0$.
The time $t_{\rm R_{max}}$ calculated using the approximate solution (\ref{tmax}) is 
 $9.4 $ days, while the observed time when $R$ is maximum is 14.6 days.
 The seismicity rate is proportional to the stress rate for $t>t_{\rm R_{max}}$.}
\end{figure}

At short times, the first term dominates in (\ref{thpl}).
Using this approximation, the friction law (\ref{mupl}) becomes 
\be
\left[ A   + { B V_0  t^* \over D_c}   - {k   V_0 t^* \over \sigma } \right ] \ln \left(1+{ t \over t^*}\right) 
=  \mbox {constant} \,,
\ee
which implies that the characteristic time $t^*$ in  (\ref{Vpl})  obeys 
\be
 t^* = {A \over k V_0 /\sigma - B V_0 / D_c}= {D_c A / V_0 \over (B-A) k/ k_c - B },
\label{ts1}
\ee
with $k_c = (B-A) \sigma / D_c$.
Depending on  $B/A$ and $k/k_c$, $q$ and $t^*$ can be either positive or negative. 
The expression of $q$ as a function of $B/A$ and $k/|k_c|$ is thus
\be
 q = {V_0 t^* \over D_c} = {A \over (B-A) k/ k_c - B }\,,
\label{q}
\ee
and is shown in Figure \ref{figq}.
$q$ is always negative in the unstable regime $B>A$ and $k<k_c$, but 
can be either positive or negative if $B<A$ or $k>|k_c|$.

When $q>0$, the approximation $K (t^* + t) ^{-q} \ll (t^* +t )/( q +1 ) $ in  (\ref{thpl}) will hold as long as 
$ t \ll  t_1 \left( {D_c /\theta_0 V_0 (q+1)} \right) ^{-1/(1+q)}$.
This corresponds to regime \#1 (see Table 1 for a list of regimes and
approximations).
If the system is initially close to steady state, i.e., $ \dot \theta_0 \approx 0$, then $ \theta_0 \approx D_c /V_0 = t^*/q $.
The transition to the next regime occurs at times of the order of $\theta_0$, when the slip is $\delta \approx D_c q \ln ( 1 + 1/q) \approx D_c$.
The slip generated during this phase thus reaches a maximum value of the order of $D_c$ 
before the transition to the next regime when the second term in (\ref{thpl}) dominates, which is regime \#4 in Table 1.

If $q<0$, then $t^*<0$ and equation (\ref{Vpl}) describes a power-law singularity of slip rate (earthquake).
In this case, we recover the solution derived by  \citep{dieterich94} for earthquake nucleation in the
absence of external loading ($\dot  \tau_r=0$).
This solution is valid only for large stress $\mu(V) >\mu_a(V)$, i.e., when slip rate initially increases with time.
This corresponds to regime \#2 in Table 1.
In addition, the system will not reach instability if $\ddot \theta >0$, corresponding to $\mu(V) < \mu_l(V)$.
If   $\mu(V) < \mu_l(V)$,  fault healing  accelerates, eventually forcing slip rate to decrease with time.
This corresponds with the slow earthquake case, discussed in
Section \ref{sloweq},
of regime \#2 transforming to regime \#6, or regime \#3 transforming
to regime \#4 or regime \#6; see Figure \ref{figq}.

The second term   $ (t^* +t )/( q +1)$ dominates at large times if $q>0$ and $t^*>0$.
Putting this approximate solution   $ \theta = (t^* +t )/( q +1)$ in (\ref{mupl}), this gives
\be
\left[ A -B - {k \over \sigma}  V_0 t^* \right ] \ln  \left(1+{t\over t^*}\right) =  \mbox {constant} \,,
\ee
which implies
 \be
t^* = {(A-B) \sigma \over k V_0} \,. 
\label{ts2}
\ee
This characteristic time is the same as found by \citet{scholz90} using the steady state approximation.
The friction coefficient given by   (\ref{ratestate},\ref{thpl}) is  
$\mu = \mu_0+ (A-B) \ln (V/V_0)$, equal to the steady-state value. 
However our solution (\ref{Vpl},\ref{thpl}) is not at steady state; healing rate given 
by (\ref{dtpl}) is
 \be
 \dot \theta = {1 \over 1 + t^* V_0 /D_c}  =  {1 \over 1  - k_c/k} \,.
\label{dthl}
\ee
This corresponds to regime \#4.

We also found another regime for which slip rate decays as a power of time, with an exponent $B/A$.
This regime characterizes afterslip when $\delta \ll A \sigma /k$, i.e., when change in friction due to slip is negligible. 
With this approximation, slip rate (\ref{V}) can be rewritten as 
\be
 V = V_0\, {\left(  \theta  \over \theta_0 \right)}^{-B/A} \,
 \label{V2}
 \ee
 and the evolution law (\ref{ratestate2})  becomes
 \be
 \dot \theta = 1 -  {V_0 \theta _0 \over D_c}  \left( { \theta \over \theta _0 } \right)^{1-B/A} 
\ee
This solution has no analytical solution. 
To obtain simple solutions, we further assume that either (i) 
$ B\approx A$, so that   $\dot \theta \approx 1 -  V_0 \theta _0 / D_c $, or (ii) slip rate 
is small, so that  $\dot \theta \approx 1$.
State variable is thus given by $\theta= \theta_0 + \dot \theta_0 t$, and slip rate 
obeys 
\be
 V =  { V_0 \over  \left(   1 +  \dot \theta_0 t /  \theta_0 \right)^{-B/A}} \,.
 \label{V3}
 \ee
Our solution  (\ref{V3}) can be applied as well in the velocity weakening case,
corresponding to regime \#5, or velocity strengthening case,
corresponding to regime \#6.
In the stable regime, the solution (\ref{V3}) will be followed by the other power law decay 
described by (\ref{Vpl},\ref{ts2}), when slip becomes large compared with $A\sigma /k$.

{}

\end{article}

\clearpage

\begin{table}
\label{tab1}
\caption{Analytical approximative solutions for afterslip}
 \begin{flushleft}
\begin{tabular}{|c | c | c | c| c|c|c|}
\tableline
		&  $1$&	$2$ &	 $3$		& $4$	& $5$ & $6$	\\
\tableline

if		 &  $q>0$ &     $-1<q<0$  &	$q<-1$	&	 $A>B$		&	 $A>B$		& $B>A$	\\
		&	 \multicolumn{3}{|c|}{	$q = {V_0\, t_1^* \over D_c} = {A  \over (B-A) k/ k_c - B } $ }&	
		&	 \multicolumn{2}{|c|}{}		\\

approx	&	 \multicolumn{3}{|c|}{	$ \dot \theta \ll 1/(q+1) $ }    &  $\dot\theta \approx \dot\theta_l $ 		
		 &  \multicolumn{2}{|c|}{  $ \delta\ll A\sigma/k $  and $\dot \theta  \approx \dot \theta_0  $ }	\\
		 &$t_1^*>0$ &  $t_1^*<0$ &  $t_1^*<0$ &		 &&	 \\
		 & $V\searrow$,  $\dot\theta\nearrow$   
		 &  $V\nearrow$,  $\dot\theta\searrow$
		 & $V\nearrow$,  $\dot\theta\nearrow$ 
		 & $V\searrow$,  $\dot\theta\rightarrow$ 
		 		  & $V\searrow$,  $\dot\theta\searrow$  & $V\searrow$,  $\dot\theta\nearrow$   \\
 $t^*$         &  \multicolumn{3}{|c|}{$ t_1^* = {A \over k V_0 /\sigma - B V_0 / D_c} =  {D_c A / V_0 \over (B-A) k/ k_c - B }$}
 		&	$ t_2^* = {(A- B)\sigma \over k V_0} = {D_c | k_c | \over k V_0}   $ 
 		&  \multicolumn{2}{|c|}{$ t_3^* = { \theta_0 \over  \dot \theta_0 }
		=\left( { 1 \over  \dot \theta_0 } -1\right) {D_c \over V_0 }  $} 			 \\
$\delta$	 &   \multicolumn{3}{|c|}{ $  q D_c \ln(1+t/t_1^*)$}  		&   $  {D_c |k_c| \over k }  \ln(1+ t / t_2^*)$ 
		&  \multicolumn{2}{|c|}{${  D_c  \over(1- B/A)} \left ( { 1 \over \dot \theta_0} -1 \right)
		  \left[(1+ {t \over  t_3^*})^{1- {B\over A}}-1\right]$}	 \\
 $V$	          &  \multicolumn{3}{|c|}{ $ {V_0 \over 1+ t/t_1^*}$   }	 &  $ {V_0 \over 1+ t/t_2^*}$
 		&   \multicolumn{2}{|c|}{$ {V_0 \over (1+  t/  t_3^*)^{B/A}}$} \\
		
 $\theta$    &   \multicolumn{3}{|c|}{ 	  $ \theta_0 (1+ t/ t_1^*)^{-q}  $}	&$ \theta_0 +\dot \theta_l t  $	
 		&  \multicolumn{2}{|c|}{$ \theta_0 +  \dot \theta_0 t$ } 	\\

 $\dot \theta$	&    \multicolumn{3}{|c|}{$1-{  \theta_0 V_0 \over D_c} (1+  { t \over  t_1^*})^{-q -1} $}	
			 &	$ \dot\theta_l={1\over 1+ |k_c|/k}$	
		 	& \multicolumn{2}{|c|}{ $ 1 - { \theta_0 V_0\over D_c} (1+ { t \over   t_3^*})^{1-B/A}$}	\\
 $\mu$	&   \multicolumn{3}{|c|}{ $ \mu_0 + (A + qB)\ln(V/V_0)$ }	 & $ \mu_0 + (A-B)\ln(V/V_0)$ &	  \multicolumn{2}{|c|}{$\mu_0$}  	 \\
 \tableline
\end{tabular}
\end{flushleft}
\end{table}

\begin{table}
\label{tab2}
\caption{Fit of afterslip data with the rate-and-state friction law. Time units is days, and displacements are expressed in m.}
 \begin{flushleft}
\begin{tabular}{|c|c|c|c|c|c|c|c|c|c|}
\tableline
data   &  $ t_{\rm min}$ & $t_{\rm max}$ & $t_2^*\tablenotemark{a}$  & $\delta_{\rm max}$ &  $p$ \tablenotemark{b} 
&  $A>B$ \tablenotemark{c}  &  $B>A$  \tablenotemark{c} 
& $B>A$ \& $k<k_c$  \tablenotemark{c} 	& $N_g/N$ \tablenotemark{d}  \\
\tableline
$N_{h,S}$    & 1& 1610 & 1.3 & 0.14 &0.92&  *& & &     76/229\\	
$N_{h,W}$  &  1& 1610 & 1.0 & 0.16 &0.88& *& & &    45/100\\ 
$ NC_{h,S}$ &   1  &539&  1.5  &0.36& 1.07 &  *& *& *& 37/100 \\
$NC_{h,W}$    &1& 1610 & 1.6&  0.41 &1.08 &* &* &*& 24/100\\
$ SC_{h,S}$    &1 &1610  &1.9  &0.48 &1.07 &* &*& *& 24/100\\
$ SC_{h,W}$    &1 &1610  &1.7  &0.48 &1.07 &* &* &* &25/100\\
$ SWC_{h,S}$   &1 &1610  &1.8&  0.54 &1.08&  *& *& *& 22/100\\
$SWC_{h,W}$ &  1 & 1610  &  2&  0.54& 1.07&*&*& *& 17/100\\
$FSSC_{h,BB}$ & 1 &1233 & 2.2 &  0.46 &1.13&  *& *&&   19/285	\\
$FSSC_{h,S}$  & 1& 1610 & 2.2 & 0.51& 1.15&   * & & &	17/200\\
$FSSC_{h,W}$  & 1& 1610&  5.8 & 0.52 &1.15&    *& *& *& 26/200\\ 
$S_{h,S}$   &  1 &1610 & 1.3 & 0.14& 1.19& *& *& *& 36/100\\
$S_{h,W}$   &  1 &1610&  1.3&  0.14& 1.18 &*&*&* &44/100\\ 
$SC_{z,S}$   & 1 &1610 & 0.63 &0.06 &1.02 &* &* &* &58/100 \\
$SC_{z,W}$   & 1 &1610 & 0.74& 0.05 &1.04& *&*&* &69/100\\ 
$S_{z,S}$  &   1 &1610 &  4.2& 0.06& 1.29& * &* &* &44/100\\
$S_{z,W}$    & 1& 1080  & 3.4 &0.06 &1.20& *& *& *& 52/100 \\
${\rm creep}_h$ & 0.0007 &0.12 &0.16 &0.005 &?\tablenotemark{e} &  &*&&  27/300	\\
\tableline
\end{tabular}

\tablenotetext{}{}		
\tablenotetext{a}{value of $t_2^*=(A- B)\sigma / k V_0$ estimated by \citet{wennerberg97}}
\tablenotetext{b}{exponent $ p=d\ln V / d\ln t$ estimated for $5t_2^* <t <t_{\rm max}$}
 \tablenotetext{c}{a star indicates there are models in this range of model parameters that fit the data. 
 with a residue smaller than ${\rm rms}_{\rm min}$.} 
  \tablenotetext{d}{Ratio of the number of models with ${\rm rms}<{\rm rms}_{\rm min}$  and of the number 
  of optimizations with different initial values of model parameters.} 
 \tablenotetext{e}{$p$ cannot be estimated because $t_2^* >t_{\rm max}$}
\end{flushleft}
\end{table}

\clearpage

\end{document}